\documentstyle[12pt]{article}
\input{psfig}
\newlength{\extraspace}
\newlength{\extraspaces}
\newcommand{\beq}{\begin{eqnarray}
\addtolength{\abovedisplayskip}{\extraspaces}
\addtolength{\belowdisplayskip}{\extraspaces}
\addtolength{\abovedisplayshortskip}{\extraspace}
\addtolength{\belowdisplayshortskip}{\extraspace}}
\newcommand{\eeq}{\end{eqnarray}}
\newcommand{\bea}{\begin{eqnarray}
\addtolength{\abovedisplayskip}{\extraspaces}
\addtolength{\belowdisplayskip}{\extraspaces}
\addtolength{\abovedisplayshortskip}{\extraspace}
\addtolength{\belowdisplayshortskip}{\extraspace}}
\newcommand{\eea}{\end{eqnarray}}

\newcommand{\goto}{\rightarrow}

\newcommand{\al}{\alpha}

\newcommand{\om}{\omega}
\newcommand{\xvec}{{\bf x}}

\newcommand{\kvec}{{\bf k}}

\begin{document}
\addtolength{\baselineskip}{.8mm}
\thispagestyle{empty}

\begin{flushright}
OU-TP-98-08P\\
{ hep-th/9802102}\\
{ February 1998}
\end{flushright}
\vspace{0.5cm}
\begin{center}

{\large\sc{VACUUM INSTABILITY IN CHERN-SIMONS THEORY, NULL VECTORS
   AND TWO-DIMENSIONAL LOGARITHMIC OPERATORS}  }\\[15mm].

{Ian I. Kogan\footnote{e-mail:
i.kogan1@physics.oxford.ac.uk} and  Alex
Lewis\footnote{e-mail:
a.lewis1@physics.oxford.ac.uk} \\}
\vspace{0.5cm}
{\it   Department of Physics, Theoretical Physics \\
  University of Oxford, 
1 Keble Road, Oxford, OX1 3NP \\ United Kingdom}\\
\vspace{0.5cm}
{\sc Abstract}
\end{center}
\noindent
{A new relation between two-dimensional conformal field theories and
three-dimensional  topologically massive gauge theories is  found,
where the dynamical nature of the 3d theory is ultimately important.
 It is shown that the  those primary states in CFT which have
non-unitary descendants correspond in the 3d theory to supercritical
charges and cause vacuum instability. It is also shown that
logarithmic operators separating the unitary sector from a non-unitary one
 correspond to an exact zero energy ground state in which case 
the 3d
Hamiltonian naturally has a Jordan structure.   }
\vfill

\newpage
\pagestyle{plain}
\setcounter{page}{1}

It is  well known that a topological Chern-Simons theory 
 on a
3-dimensional manifold with a boundary induces a WZNW model 
 on the boundary \cite{witten} which is a basic ``building block''
 for  all known unitary rational conformal field theories (CFT).
  Combining several
Chern-Simons fields and/or factorising over some discrete symmetries
 one can  give a three-dimensional construction for 
 all known unitary rational CFT \cite{MS}, for example,
minimal models \cite{bpz} through a GKO coset construction \cite{gko}.
In unitary WZNW and minimal models, primary fields only 
exist for a restricted number of
representations.  For example for  the   $SU(2)$ model,
  only the representations
with $j = 0,\frac12,\dots,\frac{k}{2}$ are allowed, while in the
minimal model the allowed  primary fields are those which satisfy the
above condition for each of the three $SU(2)$ factors in the GKO
construction. It is an interesting question why does this truncation of
 spectrum  occur from a three-dimensional point of view. 

The reason why there  must be no 
  states with $j > k/2$ is because they are physically 
 inadmissible.  Being 
themselves states with positive norm $<j|j> = \left||j>\right|^2  >
 0$ their descendants are
 ghosts - they have negative norm, so the highest weight
representation
of the current (or Virasoro algebra) is non-unitary, which is
inadmissible. 
To see this
let us calculate a norm of  a descendant
 state $J^{+}_{-1}|j>$ where the primary  highest weight  state
   $|j>$ is defined as
\beq
J^{a}_{n}|j> = 0, a =1,2,3;~ n >0; ~~~~~
J^{+}_{0}|j> =0,  ~~~~~~ J^{3}_{0}|j> = +j|j>
\label{primaryj}
\eeq
Using  a commutation relation for $SU(2)$  current algebra
\beq
[J^{-}_{n},J^{+}_{m}] = \frac{nk}{2} \delta_{n,-m} - J^{3}_{n+m}
\label{su2}
\eeq
it is easy to see that the norm of a descendant
\beq
|J^{+}_{-1}|j>|^2 = <j|J^{-}_{1}J^{+}_{-1}|j> 
 =  <j|\left(\frac{k}{2} - J^{3}_{0}\right)|j> = 
\left(\frac{k}{2}-j \right)<j|j>
\eeq
is negative provided $j > k/2$.  This negative norm is entirely due
 to the nonabelian nature of the $SU(2)$ group because the
negative contribution
 came from the second term in (\ref{su2}), which is absent  for any
 abelian current algebra. In addition there is a ``null state,'' 
$J^{+}_{-1}|k/2>$ with
zero norm. This is one of the  singular vectors of the form 
$(J^{+}_{-1})^{k-2j+1}
|j>$ which exist for any $j \leq k/2$
and which play an essential role in decoupling the non-unitary
from the unitary states \cite{gw}.

 One can ask immediately
  the following questions:
\begin{itemize}
\item {\sl  What is wrong  with  a  $2+1$-dimensional topological
CS theory if we add a source with $j > k/2$ ?}
\item  {\sl What  is so special in the so-called null states,
 i.e. states with a zero norm, from the  $2+1$-dimensional point of
view ?}
\end{itemize}

 This letter is an attempt to give answers to both of these questions
 and in one phrase one can say that the deep physical reason why
 there are no states with $j > k/2$ in a three-dimensional
description is  because they are {\it  supercritical, i.e. cause
 vacuum instability}. The answer to the second question is also quite
 amusing and it turns out 
that the null states describe the critical state
 of a three-dimensional vacuum on a verge of instability, i.e.
the  appearance of a zero-energy level. 
Remarkably, it was shown a long
time ago in \cite{ss} that  the Hilbert space of
a field theory in which this occurs has an indefinite metric and a
Jordan block structure, which we now know are characteristic features
of logarithmic conformal field theory (LCFT) \cite{gurarie}, moreover
 the ground state is two-fold degenerate and one of these states has a
 zero norm - precisely  what  the norm of one of the
  logarithmic  states in LCFT must be \cite{ckt}.

 Let us  now address the first question and  notice that a topological
  Chern-Simons theory  is a low-energy limit of a topologically massive 
 gauge theory \cite{ziegel}-\cite{tmgt}  with  the action:
\bea
S_k &=& -\frac{1}{2e^2} \int_{\cal
M}d^3x~\mbox{tr}~F_{\mu\nu}F^{\mu\nu}+ kS_{CS}
\label{action}\\
S_{CS} &=& \int_{\cal M} {1\over4\pi}~\mbox{tr}\left(A\wedge dA +
{2\over3}A\wedge A\wedge A\right)=\int_{\cal
M}d^3x~\frac{1}{4\pi}\epsilon^{\mu\nu\lambda}~\mbox{tr}\left(A_\mu
\partial_\nu
A_\lambda+\frac{2}{3}A_\mu A_\nu A_\lambda\right)
\nonumber\eea
Where $A_\mu=A_\mu^at^a$, and $t^a$ are the generators of the gauge
group $G$ which is in our case $SU(2)$.
The propagating degrees of freedom of this action are the gauge bosons
with topological mass 
$M=ke^2/4\pi$. If the manifold ${\cal M}$ has a boundary, there is a
WZNW model on the boundary with the action
\beq
S_{WZNW}(g)={k\over8\pi}\int d^2z~\mbox{Tr}g^{-1}\partial^\mu
gg^{-1}\partial_\mu g~+~{ik\over12\pi}\int
d^3z~\mbox{Tr}g^{-1}dg\wedge
g^{-1}dg\wedge g^{-1}dg.\label{wznw}\eeq 

 Actually  a TMGT with dynamical degrees of freedom 
  induces  on a boundary a deformed WZNW model  because 
massive vector particles  are charged and  it is known
  (see \cite{minmodel} and
 references therein) that charged matter in a bulk  leads to a
 deformed CFT on a boundary. This explains the difference between
abelian and non-abelian TMGT which was  discussed in \cite{acf},
 where it was shown that contrary to the  abelian case where the presence
of the Maxwell term does not lead to any deformation of the
induced conformal
 field theory on a boundary, in a nonabelian case the CFT  will be
deformed with a deformation parameter proportional to $1/e^2$.
 Thus WZNW model can be induced from a TMGT with a  boson  mass of the  order
 of the UV cut-off, so the $F^2$ term regularizes the Chern-Simons
action and we recover the WZNW model by letting the mass $M\goto \infty$ 

Although it seems  at first sight that these infinitely heavy particles 
are of no importance for low-energy dynamics, 
let us recall now that  a primary field in the representation $j$
  on the boundary
corresponds to a charged particle in the same representation in the
bulk, i.e. to a static  $SU(2)$ source in a representation $j$.
 Of course the physical vacuum  will be affected by this source, and
we might expect there to be bound states of charged gluons.
If we consider only states with a single
gluon,
we see that for small
 charges $j$ nothing is going to happen and even if the bound state
  forms its energy is still large. But when the charge $j$  increases
 the energy is going to be smaller and smaller until  at some critical value
 the vacuum instability occurs - states with negative norm will appear
  after a critical bound state with zero energy (and as we shall see
later with zero norm) is   formed. Clearly when a
heavy particle has a bound state with zero energy it can no
longer be ignored in the low-energy dynamics even if the free particle
has infinite energy. In what follows we are able to analyze 
energies for bound states of a single gluon and a charged particle. 
We expect that  there are
analogous results for bound states of several charged gluons, 
which correspond to the singular vectors at level $k-2j+1$ for the
primary states $|j\rangle$ in the WZNW model, but this is a many body
problem which we cannot solve at this stage.

One way of understanding the singular
vectors $|C\rangle \equiv (J^{+}_{-1})^{k-2j+1}
|j\rangle$
is that as well as being  descendants of primary states $|j\rangle$
they can be considered to be  primary states, as 
$J^a_n|C\rangle=0$ for $n\geq 0$. In the CS theory, we can see that a
state consisting of a charge $j$ and a gluon can be in the same
representation of $SU(2)$ as a charge $j+1$ but in general will not
have the same induced spin. The induced spin is just the conformal
dimension $j(j+1)/(k+2)$ for a single charged particle  (for details
see \cite{acks}), and the gluon
has spin $1$, so the descendant has spin $1+j(j+1)/(k+2)$ which is in
general not equal to the spin of a single particle of charge
$j+1$. When these spins are equal, which is when there is a null state
in the WZNW model at $j=k/2$, all the quantum numbers (spin and $SU(2)$ charge)
of the particle with charge $k/2+1$ are the same as that of the state
made from a charge $k/2$ and a charged gluon, so we can consider these
states to be the same. We therefore have to understand why the state
containing a particle of charge $j=k/2+1$ should have a zero norm and
decouple from the physical spectrum, and our conjecture is that this
is where critical bound states with zero energy will appear.  In what
follows we will also be looking at $N=1$ supersymmetric TMGT
\cite{ziegel},
 which
induces a supersymmetric WZNW (SWZNW) model on the boundary. The bosonic part
of a SWZNW model at level $k$ is an ordinary WZNW model at level
$k-2$ \cite{vkpr}, so we expect 
the same critical behaviour in the SUSY case, but
at $j=(k-2)/2+1=k/2$. The shift $k \goto k-2$ in $3d$ was 
discussed in \cite{acks}.

Let us now  show  that this instability indeed takes place  at $j =
k/2+1$, or $j=k/2$ in the SUSY case.
 
In the presence of a source in the representation $j$
of $SU(2)$ with the eigenvalue of $t^3$, $m=\pm j$ at the origin,
 the classical
background field is given by
\bea
&&A_\mu^\pm=0,~~~~~~~
F_{\mu\nu} \equiv \partial_\mu A_\nu^3-\partial_\nu A_\mu^3 \nonumber \\
&&\frac{1}{2}\epsilon^{\mu\nu\lambda}F_{\nu\lambda}
+ \frac{1}{M} \partial\nu F^{\nu\mu} +
\frac{2\pi m}{k}\eta^{0\mu}\delta^2(x) = 0
\eea
Which has the solution (with $A_\mu \equiv A_\mu^3$)
\beq
A_0(r) = qMK_0(Mr),~~~~~A_r(0) = 0,~~~~~
A_\theta(r)= -qMK_1(Mr)+\frac{q}{r},
~~~~~q=\frac{m}{k}
\label{bground}\eeq
$K_0(x)$ and $K_1(x)$ are the modified Bessel functions \cite{gr},
defined by
\beq
K_n(x)=\frac{\pi^{1/2}}{(n-\frac12)!}\left(\frac{x}{2}\right)^n
\int_1^\infty e^{-xz}(z^2-1)^{n-1/2}dz
\label{bessel}\eeq
>From the discussion above, we expect that the charged gauge
bosons in this background
will have critical behaviour when $q$ is large enough, and so we
consider the effective action for the  charged bosons in the
background field, which, from
eqs. (\ref{action},\ref{bground}) is 
\beq
S_{eff} = 2k\int d^3x \left[
\epsilon^{\mu\nu\lambda}A^a_\mu D_\nu A^a_\lambda
+\frac{1}{M}A^a_\mu D^2 A^{a\mu}
-\frac{1}{M}A^a_\mu D^\nu D^\mu  A^a_\nu
+\frac{1}{M}f^{a3c}A^a_\mu F^{\mu\nu} A^c_\nu \right]
\eeq
where the covariant derivative 
$D_\mu A^a_\nu = \partial_\mu A^a_\nu + f^{a3c}A_\mu A^c_\nu$. 
The equations of motion for the charged bosons are then
\bea
\tilde{F}^\pm_{\mu\nu} = D_\mu A^\pm_\nu - D_\nu A^\pm_\mu \nonumber \\
\frac{1}{2}\epsilon^{\mu\nu\lambda}\tilde{F}^\pm_{\nu\lambda}
+\frac{1}{M}D_\nu \tilde{F}^\pm_{\nu\mu}
\pm \frac{1}{M}F^{\mu\lambda}A^\pm_\lambda
\label{toohard}\eea
Rather than attempt to solve this set of three coupled 2nd order 
equations, we note that eq. (\ref{toohard}) applies unchanged to the
charged bosons in the SUSY TMGT. It is also clear that
 due to  the supersymmetry if there is a  bound state of a charged
 massive gluon with a static source there is a bound state with a charged
 gluino with the same energy and vice versa.  
 In the latter case, the equations of
motion for the fermions in the given background are simply the Dirac
equation:
\beq  
\left( i\gamma^\mu D_\mu -M \right)
\left(\begin{array}{c}\psi_1 \\ \psi_2 \end{array}\right) =0
\eeq
 which is only a pair of two first order differential equations.
A Majorana representation for the  gamma matrices is
 $\gamma_0=\sigma_2,\gamma_1=i\sigma_1,\gamma_2=i\sigma_3$.
If we make the gauge transformation
\beq
\psi \goto \psi' = \psi \exp\left[-\frac{A_0(r)}{M} - q \log(r) 
\right]
\eeq
and define, for a solution with angular momentum $m$ and energy $E$
\bea
\psi_\pm &=& \psi'_1 \pm i \psi'_2 \nonumber \\
x_\pm  &=& x_1 \pm i x_2 \nonumber \\
f(r) &=& -i x_+^{-m} \psi_+ e^{iE x_0}\nonumber \\
g(r) &=& r x_+^{-(m+1)} \psi_- e^{iE x_0}
\eea
we obtain the following equations (with $x=Mr$, $\om=E/M$)
\bea
\left(1 - \om - q K_0(x) \right) g(x) + f'(x) &=& 0 \nonumber \\
\left(1 + \om + q K_0(x) \right) f(x) + 
\left( (2q+2m+1)x^{-1} -2qK_1(x) \right)g(x) +g'(x) &=& 0
\label{fgeqns}\eea
where the modified 
Bessel functions $K_n(x)$ are defined in eq. (\ref{bessel}).
In these equations for the SUSY TMGT, we have $q=j/k$, but since we
know that in the SUSY theory $k$ is shifted by $2$ compared to the
ordinary TMGT or WZNW model, we expect the same spectrum to apply to
charged gluons in the ordinary TMGT with $q=j/(k+2)$.
Since these equations depend only on the dimensionless variables $x$
and $q$, and not on $M$ or $k$, we can immediately see that spectrum of
possible values of $\om$ is independent of $M$
(this is important as it allows us to take the limit $k=0$ later,
which is necessary to describe the $c=-2$ models as well as the WZNW
model at $k=0$). Although any bound state with $\om>0$ will be
unimportant when $M$ is infinitely large, a state with $\om=0$ will
survive.

For $x<<(1\pm\om)$, $K_0(x) \sim -\log(x)$, and $K_1(x)\sim 1/x$, so
for very small $x$ eq. (\ref{fgeqns}) become
\bea
\left\{1-\om -q\gamma +q\log(x)\right\}g +f' &\approx& 0 \nonumber \\
\left\{1+\om +q\gamma -q\log(x)\right\}f + (2m+1)x^{-1}g +g' &\approx& 0
\nonumber \\
\gamma \equiv \psi(1) +\log 2&&
\label{smallx}\eea
These are identical to the equations for a particle with angular
momentum $m$ in a Coulomb field
of a charge $q$ in $2+1$ dimensions. For $x>>1$ however, we find
\bea
(1-\om)g +f' &\approx& 0 \nonumber \\
(1+\om)f +(2(m+q)+1)x^{-1}g+g' &\approx& 0
\label{bigx}\eea
which are the equations for a free particle with angular momentum $m+q$.
Normally we would expect the lowest energy bound state to have $m=0$,
but in this case the Aharonov-Bohm effect gives an effective
centrifugal barrier even at $m=0$, while if $m$ and $q$ have opposite
signs, increasing $q$ both increases the attractive potential and
reduces the height of the centrifugal barrier.

Eq. (\ref{smallx}) leads to the following solutions for $f$ and $g$ at
small $x$:
\bea
f_m(x) = 1 + O(x^2,x^2\log x),&&m\geq0 \nonumber \\
f_m(x) = x^{-2m}\left\{ \frac{q}{2m}\log x + \frac{1-\om-q\gamma}{2m}
+\frac{q}{4m^2} + O(x^2,x^2\log x)\right\},&&m\leq -1 \nonumber \\
\label{smallfg}\\
g_m (x) = x\left\{ \frac{q}{2m+1}\log x - \frac{1+\om+q\gamma}{2(m+1)}
-\frac{q}{4(m+1)^2} + O(x^2,x^2\log x)\right\}, &&m\geq0 \nonumber \\
g_m(x) = x^{-2m-1}\left\{ 1+O(x^2,x^2\log x)\right\}, &&m\leq -1 \nonumber
\eea
Where in each case we reject the solution which is singular at $x=0$.
The solution for large $x$ of eq. (\ref{bigx}) is
\bea
f_m(x) &=& x^{-(q+m+1/2}\left\{AK_{|q+m|}(\kappa x) 
+ BI_{|q+m|}(\kappa x)\right\}
\nonumber \\
&\sim& Ae^{-\kappa x}+Be^{\kappa x} \nonumber \\
g_m(x) &=& (\om -1)f_m'(x) 
\label{bigfg}\eea
where $\kappa=\sqrt{1-\om^2}$. The constants $A$ and $B$ are
determined by the requirement $f$ and $g$ must match onto the solution
(\ref{smallfg}) at small $x$, and for a bound state to exist we must
have $B=0$. To find the ratio $B/A$ we need to know $f(x)$ and $g(x)$
in the intermediate region $x \sim 1$. We computed this numerically,
using Mathematica. We used eq. (\ref{smallfg}) for $x\leq 10^{-4}$, to
give a boundary condition for $f(10^{-4})$ and $g(10^{-4})$ 
which we used
to find a numerical solution for $f$ and $g$ in the region  $10^{-4}
\leq x \leq 8$, and matched this solution to eq. (\ref{bigfg}) for
$x\geq 8$. This is a good approximation since $K_0(8) \approx 10^{-4}
<<1$. The values of $\om$ and $q$ for which $B/A=0$ then give the
bound state spectrum. The results were that 
 for positive
$q$, there is a bound state with $m=-1$ and not with $m=0$. 
 The energy of the bound state falls very sharply when
 $q \approx 1/2$ and reaches $\om=0$ very close to $q=1/2$ (Figure 1).
 Numerically we find $\om=0$ at $q\approx 0.498$.

\begin{figure}
\vspace{-1cm}
\centerline{\psfig{figure=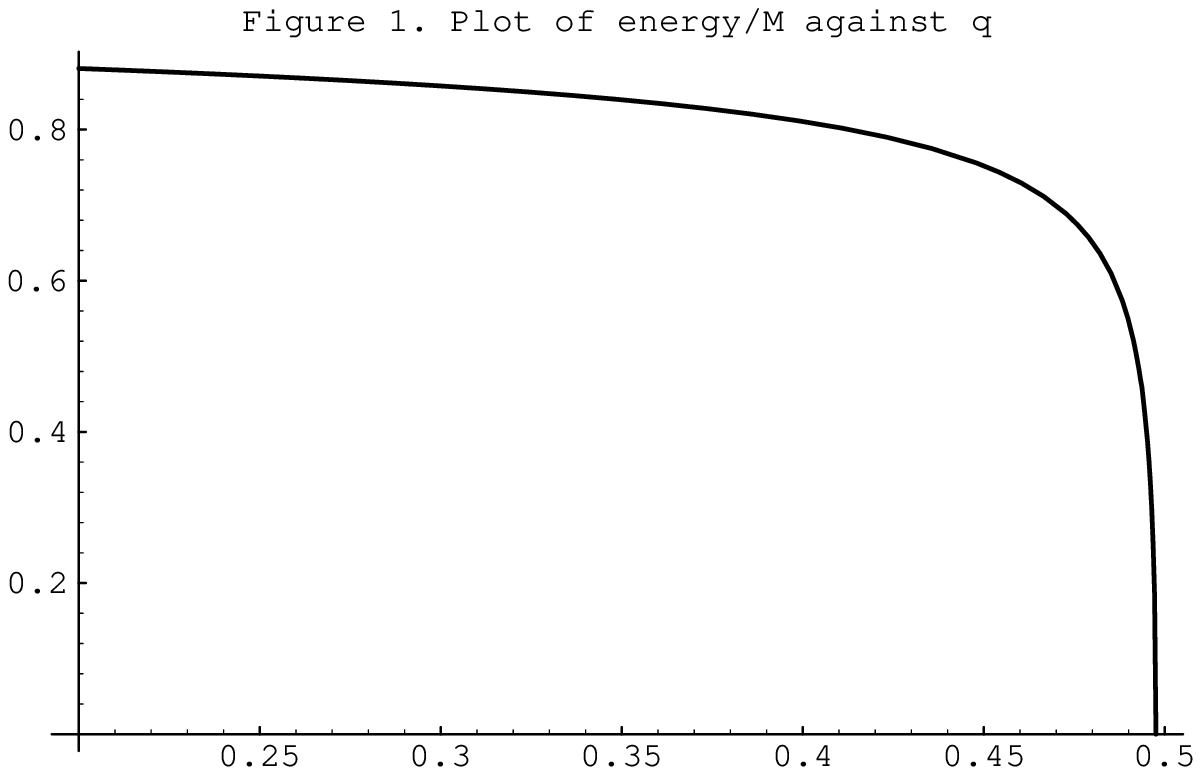,width=0.75\textwidth}}
\vspace{-10cm}
\end{figure}

  Now we  can
conjecture that in the exact solution $\om=0$ at $q=1/2$ exactly.  
When $q>1/2$, there is no bound state with real energy, and we assume
that the bound state must then have imaginary or complex energy.
This type of behaviour is not really very unusual -- precisely the
same occurs in the solution of the Dirac equation, for an
electron in the field of a point charge $Ze$  in which case  the energy
of the ground state  can be found in any textbook 
$E(Z) = m_e \sqrt{(1-\al^2Z^2)}$
so there is a critical charge $Z_c=1/\al \approx 137$, with $E(Z_c)=0$
and imaginary energy for $Z>Z_c$. The numerical solution of 
(\ref{fgeqns}) lead us to believe that the same is true for fermions
in the 
TMGT, 
with the critical charge $q_c=1/2$, and because of the supersymmetry
 the charged bosons must have the same spectrum. Because we already
 saw that   classical equations of motion for charged bosons are the
 same in supersymmetric and ordinary TMGT (apart from the shift of $k$
by $2$)
we confirmed our hypothesis
 about the  existence of a critical charge, which is $k/2$ for the
SUSY and, because of the shift $k\goto k+2$,  $k/2+1$ for the ordinary
TMGT. 
This  answers our
first question: when we add a supercritical source  to a CS theory
there are states in the spectrum with imaginary energy, leading to a
vacuum instability. There is however still an open question concerning the
charge $(k+1)/2$, which according to the above calculation appears to
be sub-critical but which, in the WZNW model, has descendants with
negative norm and does not appear in the spectrum of unitary models.

 Let us now address  the second question and study the relation between
 zero energy bound states in a bulk and null vectors and logarithmic
 operators on a boundary. The quantization of
fields with zero  energy bound states  has   been  worked
out in \cite{ss}. This was for scalar bosons, but as vector bosons in
$2+1$ dimensions only have one degree of freedom we can follow
\cite{ss} exactly.
We have the canonical
commutation relations
\bea
\left[ A({\bf x}),\Pi({\bf y})\right] = 
i\delta({\bf x}-{\bf y}) \nonumber \\
\left[ A({\bf x}),A({\bf x})\right]=0,~~\mbox{etc.}
\label{canonical}\eea
We expand the field $A$ and the conjugate momentum $\Pi$ in
terms of wave functions $\Phi^\kvec(\xvec)$ for continuum states with
$\om^2=\kvec^2+1$ and $\Phi^i(\xvec)$ for bound states with $\om_i^2 < 1$,
normalized so that
\bea
\int {\bar\Phi}^i(\xvec)\Phi^i(\xvec) d^2x=1 \nonumber \\
\int {\bar\Phi}^\kvec(\xvec) \Phi^{\kvec'}(\xvec) d^3x
=\delta(\kvec - \kvec')
\eea
We write the field operators at $t=0$ as
\bea
A(\xvec) &=& \sum_i q_i\Phi^i(\xvec) + 
\int q(\kvec)\Phi^\kvec(\xvec)d^3k \nonumber \\
\Pi(\xvec) &=& \sum_i p_i{\bar\Phi}^i(\xvec) + 
\int p(\kvec){\bar\Phi}^\kvec(\xvec)d^3k
\eea
The commutation relations (\ref{canonical}) then become 
\beq
[q_i,p_i]=i,~~~~~[q(\kvec),p(\kvec')]=i\delta(\kvec-\kvec'),~~~~~
[q_i,q_j]=0,\mbox{etc.}
\eeq
For bound states with $0<\om_i^2<1$, we introduce the mode operators
$a_i$ and $b_i$ according to a standard rule
\bea
q_i &=& \left(\frac{1}{2\om_i}\right)^{1/2}(b_i^\dagger + a_i)
\nonumber \\
p_i &=& -i\left(\frac{\om_i}{2}\right)^{1/2}(b_i - a_i^\dagger)
\label{modes}\eea
In terms of these operators, when there are no bound states with zero
or imaginary energy the Hamiltonian is
\beq
H=  \sum_i  (p_i^\dagger p_i +  \om_i^2q_i^\dagger q_i) +H_C
 =  \sum_i  (a_i^\dagger a_i + b_i^\dagger b_i) +H_C
\eeq
where $H_C$ is the part of the Hamiltonian that comes from the
continuous states. The states $|a_i\rangle \equiv a_i^\dagger
|0\rangle$
 and $|b_i\rangle \equiv b_i^\dagger
|0\rangle$ satisfy
\bea
H|a_i\rangle & =& \om_i |a_i\rangle, ~~~~~~~~~~~~~~\langle a_i | a_i
\rangle = 1  \nonumber \\
H|b_i\rangle & = &\om_i |b_i\rangle, ~~~~~~~~~~~~~~\langle b_i | b_i
 \rangle  = 1
\eea
So the Hamiltonian is of course diagonalisable and all states have
positive norm.

When there is a bound state with $\om_0=0$, its contribution 
to the Hamiltonian is just
\beq
H_0 = p_0^\dagger p_0
\eeq
In the limit $M\goto\infty$, this will be the full Hamiltonian. Now we
need to be careful about exactly how we quantize the theory. As we
will see, there are actually two choices we can make, one of which
leads to a unitary CFT on the boundary and the other to a logarithmic
CFT.
Instead of eq. (\ref{modes}), 
in this case we introduce the $p_0$ and $q_0$  as mode
operators
\beq
c=p_0,~~~~~~~d=-iq_0^\dagger
\eeq
with the commutation relations 
\beq
[c,c^\dagger]=[d,d^\dagger]=0,~~~~~~~~~[d^\dagger,c]=-1
\eeq
The Hamiltonian is $H = c^\dagger c$.  Starting from a naive vacuum
state $|0\rangle$, which of course is really a state containing a
critical charge in the full TMGT,  we can make  two more zero energy 
states,
$|c\rangle = c^\dagger |0\rangle$ and $|d\rangle =(c^\dagger +
d^\dagger) |0\rangle$. If we assume that the ``vacuum'' $|0\rangle$
has a positive norm, $\langle 0|0 \rangle=1$, we find
\bea
H|c\rangle &=& 0, ~~~~~~~~
H|d\rangle = |c\rangle \nonumber \\
\langle c| c\rangle &=& 0, ~~~~~~~~ \langle c| d\rangle =1, ~~~~~~ 
\langle d| d\rangle =2
\label{3djordan}\eea
We can see from eq. (\ref{3djordan}) that even if we give the vacuum a
positive norm, we are forced to have another eigenstate of the
Hamiltonian $|c\rangle$, which has zero energy and zero norm.
The other consistent way to quantize the theory is simply to take the
vacuum to have zero norm: $\langle 0|0 \rangle=0$. This is more
natural if we are thinking about the WZNW model as $|0\rangle$ is
actually a state with a source of charge $j=k/2+1$, which corresponds
to a singular vector in the WZNW model.
 In the latter case the state
with a charge $j=k/2+1$ would completely decouple from the physical
spectrum, just as in the WZNW model, and so this way of quantizing the
theory will give us the usual unitary WZNW model on the boundary.

The first possibility considered above, leading to
eq. (\ref{3djordan}), is what we have to  consider if we insist on adding
(super-) critical charged matter to the TMGT theory (for example, if we
want to take the SUSY model at level $2$, all non-zero charges are
excluded from the unitary theory, so this is the only 
non-trivial possibility). As was pointed out in
\cite{ss}, this does not necessarily lead to a catastrophic
instability if we allow for a Hilbert space with a metric that is not
positive definite.
As we have seen in eq. (\ref{3djordan}), the
Hamiltonian becomes  non-diagonalizable, which is a familiar property of
LCFT \cite{gurarie} where the 2d Hamiltonian, which  is a Virasoro
 operator $L_0$ acts on a logarithmic states as
\begin{eqnarray}
L_{0}|C> = \Delta |C>, ~~~~~~ L_{0}|D> = \Delta |D> + |C> \label{example}
\end{eqnarray}
where $\Delta$ is an anomalous dimension and norms of the states are
given by two-point correlation function, with one  zero norm state
\cite{ckt}
\begin{eqnarray}
\langle C(x) D(y)\rangle &&= 
\langle C(y) D(x) \rangle  = \frac{c}{(x-y)^{2\Delta_C }}\nonumber \\
\langle D(x) D(y)\rangle &&= 
 \frac{1}{(x-y)^{2\Delta_C}} \left(-2c\ln(x-y) + d\right)
\nonumber \\ 
\langle C(x) C(y)\rangle  &&= 0
\label{CC}
\end{eqnarray}

A LCFT with this behaviour exists for every central charge in the 
$c_{p,q}=1-6(p-q)^2/pq$ series, if primary fields from outside the
fundamental region of the Kac table are included \cite{f}, and
similar behaviour is expected in the WZNW model if primary fields with
$j>k/2$ are included \cite{coming}.
 It is very interesting that the Hamiltonian of  the three-dimensional
 theory in the bulk should
have exactly the same Jordan block structure as $L_0$, which is
the Hamiltonian of the two-dimensional theory on the boundary
 and in both cases we have the zero norm state.
 In the SU(2) WZNW model  $C$ and $D$  are made from the null state
 descendant of
the  $j = k/2$ state and a  $j=(k+2)/2$ state 
which have anomalous dimension
 $k/4 + 1$.  

The first primary 
field which is excluded from the unitary theory
has $j=(k+1)/2$ and  is what was called a pre-logarithmic
operator \cite{us}. Although it is not a logarithmic field, 
including this primary field in the model always
leads to the appearance of the logarithmic pair $C$ and $D$ in an
operator product expansion. This follows from the expressions for the
conformal blocks of the $SU(2)$ WZNW model given in \cite{cf}, and was
worked out in detail for the $k=0$ case 
in \cite{cklt}, and we will cover the other cases more fully in
\cite{coming}. 
The   reason
 why the prelogarithmic field with $j=(k+1)/2$ also decouples from the
spectrum is not completely clear from the $3$ dimensional point of
view. 
This is a  special case as the state $|(k+1)/2\rangle$ has
descendants with negative norm, but it is not a singular vector of
another primary state and it is 
an ordinary (not logarithmic) primary operator in a LCFT.
 One possibility is that in this case there are multi-gluon bound
states with imaginary energy, but no zero-energy state.
 We have also not
explicitly considered the effect of charges 
$j>k/2+1$, which correspond to the singular vectors
$(J^+_{-1})^{k-2j+1}|j\rangle$ for $j<k/2$. As before, this state in 3
dimensions has a source with $j<k/2$ and several gluons, but with the
same total charge and spin as a single source of charge $k-j+1$. From
the above 
calculations we would naively expect that these charges would
 be supercritical 
and the spectrum of bound states  would have complex energy 
levels, but this means that the analysis of
the energy levels is really a many body problem as these charges would
always be surrounded by negative charge gluons. We conjecture that
many body effects will lead to an exactly zero energy level for these
charges too.

The discussion above  applies with little modification to the minimal
models, and non-minimal models with the same central charges
$c_m=1-\frac{6}{m(m+1)}$. These are given by the coset $SU(2)_k \times
SU(2)_1 / SU(2)_{k+1}$ with $m=k+2$, an the three-dimensional
description is given by three $SU(2)$ fields $A,B,C$ with the action
\beq
S = S_k[A] +S_1[B] - S_{k+1}[C]
\label{minimal}\eeq
The primary fields in these models $\Phi_{r,s}$ 
have the conformal dimensions
\beq
h_{r,s}=\frac{[(m+1)r
-ms]^2-1}{4m(m+1)}
\eeq
and the field $\Phi_{r,s}$ corresponds in the three-dimensional theory
to a particle in the representations $j=(r-1)/2$ of $SU(2)_k$ and
$j'=(s-1)/2$ of $SU(2)_{k+1}$. In unitary  models only primary fields
with $1 \leq r \leq m-1$ and $ 1\leq s \leq m$ are allowed, and this
corresponds to $0 \leq j \leq k/2$ and $0 \leq j' \leq (k+1)/2$. We
can therefore see that precisely those fields which are supercritical
for at least one of the $SU(2)$ factors are excluded from the unitary
models.  It is known that if fields from outside the
unitary region are included, we get a LCFT, and the fields just
outside this region, which have $j=(k+1)/2$, are the pre-logarithmic
operators just as in the WZNW models
\cite{f,gk,rohsiepe,us}.

These calculations are good evidence that there is a critical charge
in TMGT at which a bound state of charged vector bosons (and of
fermions in the SUSY TMGT) has zero energy, which leads to a
null state in the CFT on the boundary, 
and that critical charges are related to logarithmic operators in CFT. 
Two universal features of LCFT, Jordan blocks for the
Hamiltonian and states with zero norm, also occur in the TMGT. 
Notice that the three dimensional result is equally valid in the case
of negative $k$, which is important for the coset constructions such
as
 eq. (\ref{minimal}) and the model of \cite{ctt}. 
In this letter we have only considered the TMGT and
WZNW models for $SU(2)$, but we expect similar results to apply for
other groups \cite{coming}.
Important questions which we feel deserve
 to be explored further are how the
logarithms in correlation functions of LCFT arise from TMGT, and how 
deformed LCFT is related to TMGT with  matter on the verge of
instability.

{\bf Acknowledgments}

We thank J.-S. Caux, L. Cooper, M. Flohr, G. Ross, R. Szabo,
A. Tsvelik and especially J. Cardy for helpful discussions. We are
also grateful to the referee for Phys. Lett. B for raising 
interesting questions which led us to clarify several points. The
work of A. Lewis was supported by PPARC and Brasenose College, Oxford.

\newcommand{\NPB}[1]{ Nucl. Phys. {\bf B#1}}
\newcommand{\Ann}[1]{ Ann. Phys. {\bf #1}}
\newcommand{\CMP}[1]{ Commun. Math. Phys. {\bf #1}}
\newcommand{\PLB}[1]{ Phys. Lett. {\bf B#1}}
\newcommand{\PRL}[1]{ Phys. Rev. Lett. {\bf #1}}
\newcommand{\PTP}[1]{ Prog. Theor. Phys. {\bf #1}}
\newcommand{\MPLA}[1]{ Mod. Phys. Lett. {\bf A#1}}
\newcommand{\IJMP}[1]{ Int. J. Mod. Phys. {\bf A#1}}
\newcommand{\IJMPB}[1]{ Int. J. Mod. Phys. {\bf B#1}}
\newcommand{\CQG}[1]{ Class. Quant. Grav. {\bf #1}}
\newcommand{\PRD}[1]{ Phys. Rev. {\bf D#1}}
\newcommand{\PRB}[1]{ Phys. Rev. {\bf B#1}}
\newcommand{\JMP}[1]{ J. Math. Phys. {\bf #1}}


\begin{thebibliography}{99}
\bibitem{witten} E. Witten, \CMP{121} (1989) 351.
\bibitem{MS} G. Moore and N. Seiberg, \PLB{220} (1989) 220.
\bibitem{bpz} A.A. Belavin, A.M. Polyakov and A. B. Zamolodchikov,
 Nucl. Phys. {\bf B241} (1984), 333.
\bibitem{gko} P. Goddard, A. Kent and D. Olive, \CMP{103} (1986) 105.
\bibitem{gw} D. Gepner and E. Witten \NPB{278} (1986), 493. 
\bibitem{ss} B. Schroer and J. Swieca, \PRD{2} (1970) ,2938.
\bibitem{gurarie} V. Gurarie, Nucl. Phys. {\bf B410} (1993), 535.
\bibitem{ckt} J.S. Caux, I.I. Kogan and A. Tsvelik,
Nucl. Phys. {\bf B} 466 (1996), 444.
\bibitem{ziegel}
 W.  Siegel, \NPB{156}, (1979), 135.
\bibitem{sch}
 J.F. Schonfeld, \NPB{185}, (1981),157.
\bibitem{tmgt}
 S. Deser, R. Jackiw and S. Templeton, \PRL{48} (1982)  975;
 Ann.Phys.(N.Y.) {\bf 140} (1982)  372. 
\bibitem{minmodel} I. I. Kogan, \PLB{390} (1997), 189
\bibitem{acf} M. Asorey, S. Carlip and  F. Falceto, \PLB{312}
 (1993), 477.
\bibitem{acks}  G. Amelino-Camelia, I. I. Kogan, R. J. Szabo,
\NPB{480} (1996), 413.
\bibitem{vkpr} P. Di Vecchia, G. Knizhnik, J. L. Petersen and
P. Rossi, \NPB{253} (1985), 701.
\bibitem{gr} I.S. Gradshteyn and I.M. Ryzhik,
{\it Tables of Integrals, Series and Products}, Academic Press, San
Diego, p. 1046, (1992).
\bibitem{us} I. I. Kogan and A. Lewis,  
Nucl. Phys. {\bf B} 509 (1998), 687.
\bibitem{cf} P. Christe and R. Flume, \NPB{282} (1987), 466.
\bibitem{f} M. A. Flohr, Int. J. Mod. Phys. {\bf A11}  (1996), 4147;
Int. J. Mod. Phys. {\bf A12}  (1997), 1943. 
\bibitem{gk} M. R. Gaberdiel and H. G. Kausch, Nucl. Phys. {\bf B477}
 (1996), 293; Phys. Lett. {\bf B386} (1996), 131. 
\bibitem{rohsiepe} F. Rohsiepe, hep-th/9611160.
\bibitem{ctt} J.-S. Caux, N. Taniguchi and A. M. Tsvelik, cond-mat/9801055
\bibitem{cklt} J.-S. Caux, I Kogan, A. Lewis and A. M. Tsvelik,
Nucl. Phys. {\bf B} 489 (1997), 469.
\bibitem{coming} I. Kogan and A. Lewis, work in progress 

\end{thebibliography}
\end{document}